\documentclass[reprint,amsmath,amssymb,aps,showkeys,showpacs,nofootinbib,floatfix]{revtex4-1}
\usepackage{graphicx}
\usepackage{hyperref}
\usepackage{amssymb,amsmath}
\begin{document}
\title{Test of the Weak Equivalence Principle using LIGO observations of GW150914\\and Fermi observations of GBM transient 150914}
\author{Molin Liu$^{1}$}
\thanks{Corresponding author\\E-mail address: mlliu@xynu.edu.cn}
\author{Zonghua Zhao$^{1}$}
\author{Xiaohe You$^{1}$}
\author{Jianbo Lu$^{2}$}
\author{Lixin Xu$^{3}$}
\affiliation{$^{1}$Institute for Gravitation and Astrophysics, College of Physics and Electronic Engineering, Xinyang Normal University, Xinyang, 464000, P. R. China\\
$^{2}$Department of Physics, Liaoning Normal University, Dalian, 116029, P. R. China\\
$^{3}$Institute of Theoretical Physics, School of Physics and Optoelectronic Technology, Dalian University of Technology, Dalian, 116024, P. R. China}

\begin{abstract}
About 0.4~s after the Laser Interferometer Gravitational-Wave Observatory (LIGO) detected a transient gravitational-wave (GW) signal GW150914, the Fermi Gamma-ray Burst Monitor (GBM) also found a weak electromagnetic transient (GBM transient 150914).
Time and location coincidences favor a possible association between GW150904 and GBM transient 150914.
Under this possible association, we adopt Fermi's electromagnetic (EM) localization and derive constraints on possible violations of the Weak Equivalence Principle (WEP) from the observations of two events.
Our calculations are based on four comparisons: (1)~The first is the comparison of the initial GWs detected at the two LIGO sites.
From the different polarizations of these initial GWs, we obtain a limit on any difference in the parametrized post-Newtonian (PPN) parameter $\Delta\gamma\lesssim 10^{-10}$.
(2)~The second is a comparison of GWs and possible EM waves.
Using a traditional super-Eddington accretion model for GBM transient 150914, we again obtain an upper limit $\Delta\gamma\lesssim 10^{-10}$.
Compared with previous results for photons and neutrinos, our limits are five orders of magnitude stronger than those from PeV neutrinos in blazar flares, and seven orders stronger than those from MeV neutrinos in SN1987A.
(3)~The third is a comparison of GWs with different frequencies in the range [35~Hz, 250~Hz].
(4)~The fourth is a comparison of EM waves with different energies in the range [1~keV, 10~MeV].
These last two comparisons lead to an even stronger limit, $\Delta\gamma\lesssim 10^{-8}$.
Our results highlight the potential of multi-messenger signals exploiting different emission channels to strengthen existing tests of the WEP.
\end{abstract}
\maketitle

\section{Introduction}\label{section1}

Recently, gravitational-wave (GW) event GW150914 was observed by the Laser Interferometer Gravitational-Wave Observatory (LIGO) and Virgo collaborations \cite{LIGO-PRL,LIGO-PRL2,LIGO-PRL3,LIGO-Localization,LIGO-parameters}.
Its reconstructed waveform is consistent with the signal from a binary black hole (BH) merger. Its frequency sweeps from 35~Hz to 250~Hz in duration 200~ms.
Although the sky maps have been released, including final localization from LALInference \cite{LIGO-Localization}, the precise localization is still ill-constrained.
Meanwhile, about 0.4~s after GW150914 trigger time, a weak but hard gamma-ray transient lasting 1.024~s was detected by the Gamma-ray Burst Monitor (GBM) instrument on the Fermi Gamma-ray Space Telescope, with a false alarm probability of 0.0022 \cite{VConnaughton}.
This event is known as GBM transient 150914.
Its luminosity is $1.8^{+1.5}_{-1.0} \times 10^{49}$~erg~s$^{-1}$ in hard X-ray and gamma-ray emission from 1~keV to 10~MeV.

Once the relativistic jet forms in a GW source, one may observe a prompt short gamma-ray burst (SGRB) lasting on the order of one second or less, possibly followed by X-ray, optical and radio afterglows lasting some hours or days \cite{DEichler,RNarayan}.
Due to the coincidence in their times and locations, it has been argued that GBM transient 150914 may be one such GW-triggered electromagnetic (EM) counterpart associated with GW150914 \cite{KMurase,ALoeb,RPerna,TLiu}.
In particular, a fast optical transient can emerge from a disk-driven outflow in cases where an accretion disk exists around a BH and the disk wind is driven by radiation or coronal magnetic fields in the super-Eddington accretion disk model \cite{KOhsuga,YFJiang,ASadowski,KMurase}.

Assuming that the association between gravitational and electromagnetic signals is valid, time differences between these signals may be used in principle to constrain violations of the Weak Equivalence Principle (WEP), the foundation of General Relativity (GR).
We focus here on differences in the parametrized post-Newtonian (PPN) parameter $\gamma$, which measures the curvature of spacetime around a central mass.
For one particle like a photon traveling between two given positions, propagation time in curved space will be longer than in flat space.
This time delay can be measured by radar ranging \cite{Shapiro,CWMisner}.
The first high-precision result in the Solar System came from the Viking spacecraft with an uncertainty of 0.1\%, $\gamma = 1.000 \pm 0.002$ \cite{RDReasenberg}.
Since then, various astrophysical and cosmological transients have been used to further test the WEP, such as supernovae \cite{MJLongo,LMKrauss}, gamma-ray bursts (GRBs)\cite{HGao}, fast radio bursts (FRBs) \cite{JJWei,SJTingay}, blazar flares\cite{JJWei2,ZWang} and so on.
The experimental status of the WEP is reviewed in Refs.~\cite{CMWill1,CMWill2}.

The time interval required for the travel of massless particles like photons is longer due to the presence of a gravitational potential $U(r)$ \cite{Shapiro,MJLongo,LMKrauss},
\begin{equation}\label{add-3-0-0}
\delta t = - \frac{1 + \gamma}{c^3} \int_{r_e}^{r_o} U(r) dr,
\end{equation}
where $r_e$ and $r_o$ denote the locations of source and observation from the center of potential, $c = 3 \times 10^8$~m s$^{-1}$ and $U(r)$ is the gravitational potential, which consists of three parts: the potential of Milky Way galaxy $U_{mw}$, the host galaxy $U_{host}$ and the intergalactic potential $U_{IG}$ between our galaxy and host galaxy.

Recently, the inspiral portion of the GW150914 signal with a 0.2~s time delay was used to obtain a constraint on WEP violation of order $10^{-9}$ \cite{XFWu} (see also \cite{EOKahya}).
Our study differs from Refs.~\cite{XFWu} and \cite{EOKahya} in both the method of calculation used and range of frequencies considered.
In Ref.~\cite{XFWu} the distance from the source to the center of Milky Way is equated to the distance from the source to the Earth, and in Ref.~\cite{EOKahya} the impact parameter is assumed to be at the distance ($\sim$~8~kpc) from the Sun to the Galactic center.
Because the combined LIGO and GBM observations reduce the 90\% confidence interval on position in the sky \cite{VConnaughton}, the EM localization adopted here can provide us with more definitive parameters (Sec.~\ref{section2}).
Moreover, in Refs.~\cite{XFWu} and \cite{EOKahya}, frequencies are limited from 35~Hz to the maximum amplitude frequency 150~Hz.
But multi-frequency GWs consist of signals with a bandwidth about from 35~Hz to 250~Hz, and the maximum time delay is equal to the full duration time 0.4~s (Sec.~\ref{section4}).
With this expanded frequency range, we can obtain stronger constraints.

We note also that a time difference between separate detectors was measured with high precision by LIGO.
GW150914 arrived first at Livingston ($L1$) and $6.9^{+0.5}_{-0.4}$~ms later at Hanford ($H1$), a light-travel distance of 10~ms away from $L1$ \cite{LIGO-PRL}.
Is this shift is related to gravitational time delay, and can it be used to test WEP?
Wave-particle duality tells us that the initial GWs arriving at the two LIGO detectors should be dual to a graviton as a particle with given energy.
The latter can then be used to test the WEP.

Based on the above considerations, we obtain the constraints on violations of the WEP from GW150914 and GBM transient 150914, using Fermi's possible EM localization.
Four kinds of the constraints are considered here:
First is the comparison between two detectors' initial GWs, which have different polarizations (Section~\ref{section2}).
Second is the comparison between GWs and EM waves (Section~\ref{section3}).
Third is the comparison between GWs from GW150914 in the full frequency range from 35~Hz to 250~Hz, and fourth is the comparison between EM waves from GBM transient 150914 in the full energy range from 1~keV to 10~MeV (Section~\ref{section4}).


\section{Constraints on WEP by initial GWs with minor astrophysical intrinsic time delay}\label{section2}
According to the wave-particle duality, all waves have the nature of particle and vice versa. The wave number can be interpreted as being a measure of the momentum of a particle \cite{Feynman,Greiner}. The initial GW with wavelength $\lambda_0$ or frequency $f_0$ is dual to a graviton with momentum $p_0$ or energy $E_0$ by de Broglie's equation $\lambda_0 = 2\pi\hbar/p_0$. Hence, one can test WEP by the initial graviton as being of a particle with a momentum or energy. In subsection \ref{subsection1} we present the model of calculation, which can be applied to initial GWs. This model has been well used to test WEP with GRBs and FRBs \cite{HGao,JJWei,JJWei2}. In subsection \ref{subsection-add} we discuss the difference of polarizations to ensure initial GWs can be used to constrain WEP. In subsection \ref{subsection2} we present the results of constraints on WEP by initial GWs. In subsection \ref{subsection3}, we discuss the uncertainty caused by the distance difference between separated detectors.

\subsection{The model of constraint on WEP}\label{subsection1}
In recent tests of WEP with GRBs and FRBs \cite{HGao,JJWei,JJWei2}, the time delays of observation for a cosmic transient sources contain mainly as followings,
\begin{equation}\label{add-121}
\Delta t_{obs} = \Delta t_{int} + \Delta t_{LIV} + \Delta t_{spe} + \Delta t_{gra},
\end{equation}
where $\Delta t_{int}$ is the astrophysical intrinsic time delay, $\Delta t_{LIV}$ is the potential delay by Lorentz invariance violation, $\Delta t_{spe}$ is the delay due to special relativistic effects, $\Delta t_{gra}$ is caused by the gravitational potential (for detailed explanations, see \cite{HGao}).

There are some advantages for initial outburst GWs in the respect of the uncertainty for these delays. The differences of the energies or frequencies can reach a very small value, and thus the time differences $\Delta t_{spe}$ and $\Delta t_{LIV}$ are ignorable. It should be noted that there is also no
clear evidence to indicate the frequencies of initial GWs take the same value (see the famous Fig.1 in \cite{LIGO-PRL}). The initial gravitons are most likely to be emitted simultaneously, and later propagate in the same gravitational environments including the host galaxy, the intergalactic and our galaxy. Therefore, the uncertainty of $\Delta t_{int}$ is restricted in a very small range. To this regard, the initial GWs provide us a constraint on WEP with lower uncertainty on $\Delta t_{int}$, $\Delta t_{LIV}$ and $\Delta t_{spe}$. Hence, after leaving out the negligible components $\Delta t_{int}$, $\Delta t_{LIV}$ and $\Delta t_{spe}$, the maximum delay caused by gravitational field is given by
\begin{equation}\label{1-gra-obs}
\Delta t_{gra} < \Delta t_{obs}.
\end{equation}

According to the formula (\ref{add-3-0-0}), the initial gravitons arrived at $L1$ and $H1$ have the time delays of $\delta t_{H}$ and $\delta t_{L}$ shown below due to the gravitational potential $U(r)$,
\begin{eqnarray}
\delta t_{H} &=& - \frac{1 + \gamma_{H}}{c^3} \int_{r_e}^{r_H} U(r) dr,\label{dtth1}\\
\delta t_{L} &=& - \frac{1 + \gamma_{L}}{c^3} \int_{r_e}^{r_L} U(r) dr,\label{dttl1}
\end{eqnarray}
where $r_H$ and $r_L$ are the distances from the center of potential to detectors $H1$ and $L1$. Following the original setting \cite{HGao}, the potentials $U_{host}$ and $U_{IG}$ can be ignored, and thus the dominant potential comes from $U_{mw}$ in our Milky Way galaxy. $r_H$ and $r_L$ then correspond to the distances from the center of our Milky Way galaxy to the detectors. Meanwhile, it is known that the distance of the Earth from the center of the Milky Way is about $R_{o} \sim 8~\text{kpc}$, and the maximum value of the difference $|r_H - r_L|$ is the distance 3000~km between $H1$ and $L1$. We then obtain the maximum value of $|r_H - r_L|$ relative to $R_{o}$ shown by
\begin{equation}\label{relativerate}
  \frac{|r_H - r_L|}{R_o} \lesssim 1.3 \times 10^{-14}.
\end{equation}
Obviously, when compared with $R_o$, the difference $|r_H - r_L|$ is clearly a very small value, and therefore can be ignorable. This result immediately offer us a important relationship: $r_H \approx r_L \approx R_{o}$ in our calculating. The gravitational time delay of initial gravitons arrived at $H1$ and $L1$ can be obtained by the difference between $\delta t_{H}$ (\ref{dtth1}) and $\delta t_{L}$ (\ref{dttl1}), and shown by
\begin{equation}\label{chushi}
\Delta t_{gra} = \delta t_{H} - \delta t_{L} = \frac{\Delta \gamma}{c^3} \int^{R_o}_{r_e} U(r) dr,
\end{equation}
where $\Delta \gamma = |\gamma_H - \gamma_L|$ is the difference of PPN parameters $\gamma$ for the different initial gravitons arrived at $L1$ and $H1$. We then adopt the popular Keplerian potential model, namely $U(r) = U_{mw}(r) = -GM_{mw}/r$, in which $G = 6.67 \times 10^{-11}~\text{m}^3~\text{kg}^{-1}~\text{s}^{-2}$ is the gravitational constant, $M_{mw}$ is the mass of Milky Way galaxy and is estimated as $6 \times 10^{11} M_{\odot}$ \cite{PJMcMillan,PRKafle}. Substituting above potential into Eq.(\ref{chushi}) and using the relationship (\ref{1-gra-obs}), we can obtain $\gamma$ difference of initial gravitons received by $H1$ and $L1$ as followings,
\begin{equation}\label{0-2}
|\gamma_H - \gamma_L| < \Delta t_{obs} \frac{c^3}{G M_{mw}\log (d/b)} ,
\end{equation}
where $d$ is the distance from the transient to Earth. $b$ is the impact parameter of the light rays relative to the center of the galaxy. The impact parameter $b$ is given by
\begin{equation}\label{0-3}
b^2 = r^2_G \left[1 - \left(\sin\delta_S \sin\delta_G + \cos\delta_S \cos\delta_G \cos\Delta \beta\right)^2\right],
\end{equation}
where the cosmic source is in the direction of R.A. = $\beta_S$, Dec. = $\delta_S$.
The distance from Sun to galaxy center is $r_G = 8.3~\text{kpc}$. The galaxy center is in the coordinates of ($\beta_G = 17^h 45^m 40.04^s$, $\delta_G = -29^o 00' 28.1''$) by the equatorial coordinate system J2000.0 \cite{SGillessen}. $\Delta \beta = \beta_S - \beta_G$ is the difference of R.A. between source and galaxy center.
\begin{figure}
\includegraphics[width=3.4 in]{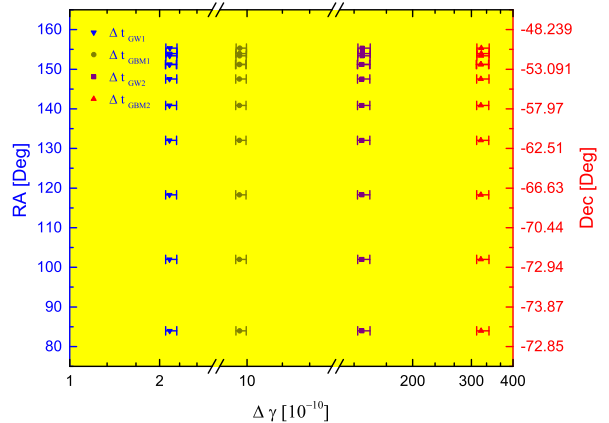}
\caption{The up-limit of $\Delta \gamma$ of initial GWs or gravitons for Fermi's 11 possible positions in the LIGO arc. The error bars come from the uncertainty of luminosity distance, and correspond to 570~Mpc and 230~Mpc from left to right. }\label{Fig1}
\end{figure}
\subsection{The different polarizations of initial GWs}\label{subsection-add}
The alternative statement of WEP is that the trajectory of a freely falling ``test" body is independent of its internal structure and composition \cite{CMWill2}. In order to ensure the initial GWs or gravitons can be used to constrain WEP, we must be clear to show that they are essentially different waves or particles.

Considering the situation of that the gravitational radiation is completely described by GWs' polarizations \cite{LIGO-parameters}, we thus need to check the polarizations of initial GWs arrived at $L1$ and $H1$. The strain $h_k(t)$, i.e. $h_k(t) = (\delta L_x - \delta L_y)/L$, measured by each instrument can be expressed by two independent and time-dependent polarizations $h_+$ and $h_\times$ through followings formulas,
\begin{equation}\label{ad_strain}
h_k = F^{(+)}_k h_+ + F^{(\times)}_k h_\times,
\end{equation}
where $F^{(+)}_k$ and $F^{(\times)}_k$ are two weighted coefficients and depend on the orientation of source, and therefore have an approximation for initial GWs: $F^{(+, \times)}_{H1} \approx F^{(+, \times)}_{L1}$. The two independent polarizations $h_+$ and $h_\times$ can be expressed as
\begin{eqnarray}
\label{h+} h_+ (t) &=& A (t) (1 + \cos^2 \iota) \cos \phi (t),\\
\label{h-} h_\times (t) &=& - 2 A (t) \cos \iota \sin \phi (t),
\end{eqnarray}
where $A (t)$ and $\phi (t)$ are the GW amplitude and phase, $\iota$ is the orbital inclination of binary. Clearly, for the initial GWs the parameter $\iota$ is also unchange. Hence, as the most conservative estimate, in the parameters space: $F^{(+, \times)}_{H1}$, $F^{(+, \times)}_{L1}$, $\iota$, $A (t)$, $\phi (t)$, at least the last two parameters are different between the initial GWs arrived at $L1$ and $H1$. About the detailed explanations of these inferred parameters, please refer to \cite{LIGO-parameters}.

According to the data released by LIGO \cite{LIGO_data}, in which the Hanford strain is shifted back in time by $6.9^{+0.5}_{-0.4}$~ms, we can find the initial strain difference $\Delta h = h_{L1} - h_{H1}$ exists between $L1$ and $H1$. Here, we ignore the difference of detectors' calibration between $L1$ and $H1$, and therefore we can get $\Delta h_k = \Delta h^M_{k}$ where $h_k$ is the physical strain appeared in Eq.(\ref{ad_strain}) and $h^M_{k}$ is the measured strain. The initial GWs or gravitons should be corresponded to the initial strains observed by $H1$ and $L1$. The released data clearly show the initial strain at $H1$ is about $2.45 \times 10^{-2}$ and the initial strain at $L1$ is about $-1.19 \times 10^{-1}$ \cite{LIGO_data}.
Therefore, the relative differences are $\Delta h/h_{L1} \approx 1.21$ for $L1$ and $\Delta h/h_{H1} \approx -5.86$ for $H1$, and can not be ignored clearly. So we can get a point of that the initial GWs or gravitons received by $L1$ and $H1$ have different polarizations.

For the initial GWs, $F^{(+, \times)}$ and $\iota$ should have the same magnitudes since these GWs are radiated simultaneously in the model hypothesis. In this way, at least one of the last two parameters $A (t)$ and $\phi (t)$ should be different in Eq.(\ref{ad_strain}) for $L1$ and $H1$. So based on above situations we can quantitatively give the differences about the polarizations.

We consider the first case which has the same amplitudes, i.e. $A_{L1} = A_{H1} = A$. According to Eqs.(\ref{ad_strain}), (\ref{h+}) and (\ref{h-}), the strains of $L1$ and $H1$ are given by
\begin{eqnarray}
\nonumber h_{L1} &=& F^{(+)} A (1 + \cos^2 \iota) \cos \phi_{L1} - 2 F^{(\times)} A \cos \iota \sin \phi_{L1},\\
\nonumber h_{H1} &=& F^{(+)} A (1 + \cos^2 \iota) \cos \phi_{H1} - 2 F^{(\times)} A \cos \iota \sin \phi_{H1}.
\end{eqnarray}
These formulas can explicitly give strains' difference as
\begin{equation}\label{ad_dstrain1}
  \Delta h = \mathcal{F}_1 \sin \frac{\Delta \phi}{2},
\end{equation}
where $\Delta \phi = \phi_{L1} - \phi_{H1}$ is the phases difference, $\mathcal{F}_1$ is the function of the sum of phases $\varphi = \phi_{L1} + \phi_{H1}$ and shown by
\begin{equation}\label{ad_dstrain12}
\mathcal{F}_1 = -2 F^{(+)} A (1 + \cos^2 \iota) \sin \frac{\varphi}{2} + 4 F^{(\times)} A \cos \iota \cos \frac{\varphi}{2}.
\end{equation}
According to the strains difference $\Delta h$ (\ref{ad_dstrain1}), the difference of the phases is found for the initial GWs. Otherwise the strains $h_{L1}$ and $h_{H1}$ are equal, which is not compatible with the observation.

Then we discuss the second case which has the same phases, i.e. $\phi_{L1} = \phi_{H1} = \phi$. According to Eqs.(\ref{ad_strain}), (\ref{h+}) and (\ref{h-}), the strains $h_{L1}$ and $h_{H1}$ are given by
\begin{eqnarray}
\nonumber h_{L1} &=& F^{(+)} A_{L1} (1 + \cos^2 \iota) \cos \phi - 2 F^{(\times)} A_{L1} \cos \iota \sin \phi,\\
\nonumber h_{H1} &=& F^{(+)} A_{H1} (1 + \cos^2 \iota) \cos \phi - 2 F^{(\times)} A_{H1} \cos \iota \sin \phi.
\end{eqnarray}
In this case, the GWs' amplitude difference $\Delta A = A_{L1} - A_{H1}$ gives the strain difference $\Delta h$ shown by
\begin{equation}\label{ad_aamp}
\Delta h = \mathcal{F}_2 \Delta A,
\end{equation}
where the function $\mathcal{F}_2$ is unchanged for initial GWs in this case and is expressed by
\begin{equation}\label{ad_aamp1}
\mathcal{F}_2 = F^{(+)} (1 + \cos^2 \iota) \cos \phi - 2 F^{(\times)} \cos \iota \sin \phi.
\end{equation}
Like the former case if the amplitudes of initial GWs arrived at $L1$ and $H1$ have the same magnitude, the strains $h_{L1}$ and $h_{H1}$ must be the same, which is also not compatible with the observation.

Until now we can know that, according to the data release for event GW150914 \cite{LIGO_data}, the initial GWs indeed are not the same, and have clearly different polarizations. Specifically, these waves have the different phase (the first case), or have the different amplitude (the second case), or have one combination of two cases. On the other hand, based on the geometric optics for GWs of small amplitude \cite{CWMisner}, the polarization tensor $\textbf{e}$ has the transverse ($\textbf{\emph{e}} \cdot \textbf{\emph{p}} = 0$) and traceless ($e_{\alpha}^{\alpha} = 0$) properties, in which $\textbf{\emph{p}}$ is 4-momentum. Thus, the gravitons with different polarizations have different 4-momentums. So the initial GWs or gravitons received by $L1$ and $H1$ have the different polarizations and 4-momentums, and therefore have fully different internal structure which ensures initial GWs or gravitons can be used to test WEP.
\subsection{The results of initial GWs}\label{subsection2}
We then use the localization of Fermi to constrain the WEP with the initial GWs or gravitons. It is known that with the $68\%$ statistical uncertainty region over 9000 square degrees ($\sigma = 54^\circ$), Fermi GBM presented a localization of best fit position (BFP) (RA = 57~deg, Dec = -22~deg) to the hard model spectrum. Substituting BFP and observed delay time $\Delta t_{GW1} \equiv 6.9^{+0.5}_{-0.4}$~ms into Eqs.(\ref{0-2}), we can obtain the up-limit of $\Delta \gamma$ as
\begin{equation}\label{1-1}
\big|\gamma_{H} - \gamma_{L}\big| < 2.13^{+0.12}_{-0.06}\times 10^{-10},
\end{equation}
which is two orders of magnitude tighter than the result $10^{-8}$ of recent $FRB110220$ \cite{JJWei} and $FRB150418$ \cite{SJTingay}.
\begin{figure}
\includegraphics[width=3.4 in]{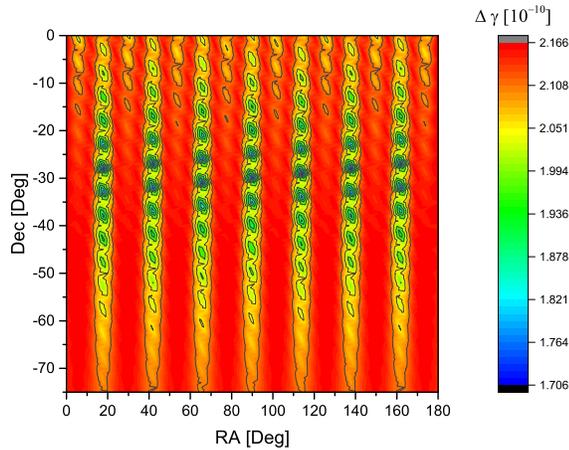}
\caption{The contour drawing of the up-limit $\Delta \gamma$ of initial GWs or gravitons with continuous R.A. and Dec. in the LIGO arc. The up-limit of $\Delta \gamma$ keeps the same order of magnitude and varies from the minimum $1.706 \times 10^{-10}$ to the maximum $2.166 \times 10^{-10}$.}\label{Fig2}
\end{figure}

Furthermore, Fermi also divides the LIGO arc into $11$ positions with $5^\circ$ apart. $10$ positions of them are on the southern portion and the last one is on the north (see Table 2 in \cite{VConnaughton}). According to the $11$ possible sky locations, we can calculate $\Delta \gamma$ with two kinds of uncertainties on distance and time. The results are illustrated in Fig.\ref{Fig1}. The average value is $2.16 \times 10^{-10}$. The uncertainty of distance $410^{+160}_{-180}$~Mpc shows that the maximum positive deviation is $1.23 \times 10^{-11}$ with $5.8\%$ accuracy and the maximum negative deviation is $6.4 \times 10^{-12}$ with $3\%$ accuracy. The uncertainty of delay time $6.9^{+0.5}_{-0.4}$~ms indicates that the maximum positive deviation is $1.57 \times 10^{-11}$ with $7.3\%$ accuracy and the maximum negative deviation is $1.25 \times 10^{-11}$ with $5.8\%$ accuracy. Obviously, the results are more easily influenced by the uncertainty of time delay.

Taken as a whole, the results tell us that the up-limits of $\Delta \gamma$ are all kept in the same order of $10^{-10}$ no matter for BFP localization or for other $11$ positions. In order to further confirm it, we calculate the whole LIGO arc with continuous angle change and obtain the contour profile of up-limit of $\Delta \gamma$ variation in the ranges of RA = $[0, 180^\circ]$ and Dec = $[-75^\circ, 0]$ in Fig.\ref{Fig2}. It illustrates that the up-limits of $\Delta \gamma$ all have the same order of magnitude $10^{-10}$ with the maximal value $2.166 \times 10^{-10}$ and the minimum value $1.706 \times 10^{-10}$.
\subsection{The influence of distance difference between separated detectors}\label{subsection3}
\begin{figure}
\includegraphics[width=2.2 in]{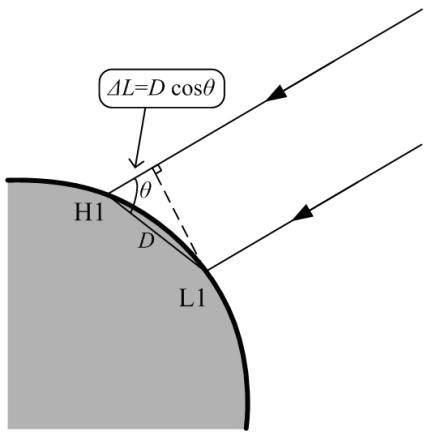}
\caption{The path difference of GWs or gravitons $\Delta L$. GWs or gravitons were propagated from the source to the detectors $H1$ and $L1$ located at the great circle on Earth. In order to ensure GWs are detected earlier at $L1$ and also to avoid the superluminal motion, the condition $D \cos \theta / c < 6.9$~ms must be respected and the angle of incidence $\theta$ should be in the range of $[-90^\circ, -46.4^\circ] \bigcup [46.4^\circ, 90^\circ]$.}
\label{Fig3}
\end{figure}

Considering the situation of that GWs arrived at separated detectors located at two different places, we need to consider the distance difference between two detectors. The actual path difference of gravitons is illustrated in Fig.\ref{Fig3} and given by
\begin{equation}\label{add-pathdiff}
\Delta L = D \cos \theta,
\end{equation}
where D is the distance of 3000~km or 10~ms light travel time between $H1$ and $L1$. Here, we define a incident $\theta$ between the incident directions and the line D connected $H1$ and $L1$. Multiplying Eq.(\ref{add-pathdiff}) on both sides by the speed of graviton which is assumed to be equal with the speed of light, we can have the graviton travel time during the path difference as $\Delta L/c = D \cos \theta /c$ where $D/c$ is the light travel time between $H1$ and $L1$. After the deduction of travel time between the detectors, the time delay caused by gravitational field satisfy the relationship
\begin{equation}\label{add-3-1}
\Delta t_{gra} < (6.9 - 10 \cos \theta)\text{~ms},
\end{equation}
where $-90^\circ < \theta < -46.4^\circ$ or $46.4^\circ < \theta < 90^\circ$. Because the present localization of GW150914 is ill-constrained, it is impossible to get a more accurate result of $\theta$ by current observation data alone. The minimal time delay takes place at the angle $\pm 46.4^\circ$ which corresponds to the situation of that 6.9~ms delay completely comes from the path difference of GWs without any gravitational effect on time delay. On the contrary, the maximum time delay occurs near at the angle of $90^\circ$ for the normal incidence of graviton beams with the same path propagated from the source, which could be adopted here to estimate the maximum gravitational time delay, i.e. $\Delta t_{GW1} = \Delta t_{L1} - \Delta t_{H1} = 6.9^{+0.5}_{-0.4}$~ms where $\Delta t_{L1} = t_{L1}(arrive) - t_{L1}(emit)$ and $\Delta t_{H1}= t_{H1}(arrive) - t_{H1}(emit)$ are the time differences of initial gravitons due to propagation from the source.

We then calculate the uncertainty caused by aforementioned path difference $\Delta L$ (\ref{add-pathdiff}).
If a perturbation $d \rightarrow d + \delta d$ along the path is considered, the up-limit of $\Delta \gamma$ (\ref{0-2}) can be rewritten in a new form of $\gamma_1 - \gamma_2 = \Delta \gamma + \delta \Delta \gamma$. So the difference caused by $\delta d$ can be expressed through $\delta \Delta \gamma$ shown below as
\begin{equation}\label{add-3-1-2}
|\delta \Delta \gamma| = \Delta t_{obs} \frac{c^3}{G M_{mw}\log^2 (d/b)} \frac{\delta d}{d}.
\end{equation}
The relative deviation is then given as followings,
\begin{equation}\label{add-3-1-3}
\Delta = \bigg|\frac{\delta \Delta \gamma}{\Delta \gamma}\bigg| = \frac{1}{\log (d/b)} \frac{\delta d}{d}.
\end{equation}
The maximum perturbation $\delta d$ comes from the distance of 6.9~ms light traveled. This point is also illustrated by the geometrical triangle relation in Fig.\ref{Fig3}. Hence, according to Eq.(\ref{add-3-1-3}), one can get $\Delta \sim 10^{-20}$ or $\delta \Delta \gamma \sim 10^{-30}$. It is clear that even for the maximum uncertainty the influence of $\delta d$ on the results is also very small, and can be ignored for the cosmic source GW150914. Actually, because the celestial equatorial coordinate system related to a cosmic source is very big for two points on Earth, it is reasonable to ignore the difference of coordinates about $L1$ and $H1$ in this large coordinate system.
\section{Constraints on WEP by GWs and EM waves}\label{section3}
Then we turn to constrain WEP by comparing GW150914 and its possible EM counterparts. Because in the real astronomical environment, GWs and EM waves would not like to radiate simultaneously. Here we consider a popular model of super-Eddington accretion for the emission modeling of EM counterpart to binary black hole merger \cite{KMurase,RPerna}, and thus get a more real astrophysical intrinsic time delay $\Delta t_{int}$ between GWs and EM waves. In this accretion model, one of two BHs is assumed being surrounded by the ``dead" fallback disk. With the temperature decreasing, the disk becomes cooled and neutral. The magneto rotational instability (MRI) will be suppressed and the disk survives in a long time as a ``dead" disk. If the BHs start to merge, the tidal torques and shocks heat up the gas. The naked BH spirals inward plowing its way within the disk. Accretion resumes from the outer regions towards the inner ones. The mass piled up propagates inwards as the inner parts of the disk get gradually revived. Following the merger, the disk is fully revived immediately. The mass piled up and accretes, powering a short Gamma-ray burst.

It is known that GWs and its possible EM waves would not like to explode at the same time, i.e. $\Delta t_{int} \neq 0$ in Eq.(\ref{add-121}). Here, $\Delta t_{int}$ can be used to quantitatively describe the exploded time delay between GWs and EM waves for the binary compacted stars as source. It is necessary to first give the range of $\Delta t_{int}$ according to the characteristics of signals, before calculating the gravitational time delay $\Delta t_{gra}$. There are two main points to be indicated. (i) According to LIGO's report \cite{LIGO-PRL} (or see the review \cite{LIGO-August}), the amplitude of GW150914 is initially increasing at the time mark 0.3s. And the possible SGRB most likely explodes in the stage of signal enhancement. Therefore, the reasonable outbreak time of possible EM companion should be later than $0.3$s, i.e. $\Delta t_{int} > 0.3$s. (ii) According to Fermi's report \cite{VConnaughton}, GBM transient 150914 was detected later than GW150914 about $0.4$s. So, we can have $\Delta t_{int} < 0.4$s. Otherwise, the speed of EM waves will be faster than that of GWs if $\Delta t_{int} > 0.4$s. Clearly, it does not agree with previous assumption of that the speeds of EM waves and GWs are all have the same value as that of light. To these regards, the intrinsic time difference between GW150914 and its EM companion should be in the range of
\begin{equation}\label{add_0304s}
0.3 s < \Delta t_{int} < 0.4 s.
\end{equation}

Under the condition (\ref{add_0304s}), a reasonable $\Delta t_{int}$ between GWs and its EM companion can be given out through a super-Eddington accretion model \cite{RPerna}. In this model, the viscous time of the disk is $t_{vis} = 1/\alpha \Omega_K (R_d/H)^2$, where $\Omega_K = (G M_{BH}/R_d^3)^{1/2}$ is the Kepler rotation frequency and $H$ is the disk scale height and $\alpha$ is the viscosity parameter in unit of $\alpha_{-1} \equiv \alpha/0.1$ \cite{KMurase,RPerna}. Meanwhile, according to GWs radiation theory of binary compacted stars, the GWs inspiral timescale $t_{GW}$ is given by \cite{CWMisner,Hughes,RPerna},
\begin{equation}\label{4th-inspiral}
t_{GW} = \frac{5}{256} \frac{c^5}{G^3} \frac{R_d^4}{2 m^3} \sim 0.37sR_8^4m_{30}^{-3},
\end{equation}
where we adopt $GM_{\odot} = 1.33 \times 10^{26}\text{cm}^3\text{s}^{-2}$, $R_8 = R/(10^8 \text{cm})$, $m_{30} = m_{BH}/(30 M_{\odot})$. Here, we use the notation $Q = 10^x Q_x$ in CGS unit. Then substituting parameters $R_8$ and $m_{30}$ into Eq.(\ref{4th-inspiral}), the range (\ref{add_0304s}) can be rewritten in a specific form as
\begin{equation}\label{2th-ad-R-m}
0.806 < R_8^4 m_{30}^{-3} < 1.075.
\end{equation}
In principle, the parameters satisfied the conditions (\ref{2th-ad-R-m}) are all allowed. Taking into account the current research status, as a general consideration, we select $R_8\sim1$ and $m_{30}\sim1$. In this way, we can get a quantitative estimate: $t_{GW} \sim 0.37 s$.

If the disk remains cold by merger time shorter than viscous time, the post-merger violent accretion will occur. The viscous time and inspiral time are essentially dynamical since the parameters of disk and BHs are dynamical with the evolution of event. If the disk is reacted and two BHs merge on a very short timescale $t_{GW}$, an electromagnetic emission will be followed on the timescale $t_{visc}$. For a relatively thin disk with $R_d/H \sim 3$ at the tidal truncation radius, the burst duration from the viscous timescale at the critical radius is $t_{GBM} = 0.05 (R_d/H)^{16/5} m_{30} \alpha_{-1}^{8/5} \text{s} \sim 0.2 \text{s}$ in which the tidal truncation radius is $R_{TT} \sim 0.3 R_d$ \cite{RPerna,Paczynski}. This result agrees with the lasting time of GBM transient 150914 \cite{VConnaughton}.

Then, the EM companion is produced after the two BHs merge on the very short timescale $t_{GW}$ (see Fig.2 in \cite{RPerna}). Therefore, the inspiral timescale $t_{GW}$ (\ref{4th-inspiral}) can be equivalent to the intrinsic time, i.e. $\Delta t_{int} \sim t_{GW} \sim 0.37 s$. After subtracting intrinsic time $\Delta t_{int}$ from the total observed time 0.4~s, we can obtain an up limit of gravitational time delay $\Delta t_{GBM1}$, i.e. $\Delta t_{GBM1} \equiv \Delta t_{gra} < 0.4~\text{s} - 0.37~\text{s} = 0.03~\text{s}$.

After obtaining the gravitational time delay, we can use the method in Section \ref{section1} to constrain WEP by GWs and its EM companion. The distance from the center of potential to the observation station (LIGO and GBM) is assumed to be still the value of $R_o$. Then using Eqs. (\ref{add-3-0-0}) and (\ref{chushi}) with $\Delta \gamma = |\gamma_{gra} - \gamma_{pho}|$, we can obtain the difference between GWs and EM waves shown by
\begin{equation}\label{2-eq1}
\big|\gamma_{gra} - \gamma_{pho}\big| < 0.03 \text{s} \frac{c^3}{G M_{mw}\log (d/b)},
\end{equation}
where the relationship $\Delta t_{gra} < 0.03~\text{s}$ is used. Substituting the coordinate of BFP into the impact parameter $b$ (\ref{0-3}), we can finally obtain an up-limit of $\Delta \gamma$ between gravitons and photons,
\begin{equation}\label{2-1}
\big|\gamma_{gra} - \gamma_{pho}\big| < 9.28^{+0.5}_{-0.3}\times 10^{-10}.
\end{equation}
The corresponding results of other Fermi's $11$ sky locations are drawn by dark yellow circle points illustrated in Fig.\ref{Fig1}. Like the case of $\Delta t_{GW1}$, they are all in the same order $10^{-10}$ of magnitude and the average value is $9.40 \times 10^{-10}$ with $5.6\%$ maximum positive deviation and $3.0\%$ maximum minus deviation. At the last, substituting the semi-major axis 6913~km of GBM \cite{orbitprare_parameter} into Eq.(\ref{add-3-1-3}), one can obtain $\Delta \sim 10^{-20}$ or $\delta \Delta \gamma \sim 10^{-30}$ for the comparison between GW150914 and GBM transient 150914. The results are the same orders of magnitude as the initial GWs case in section \ref{section2}. Like the initial GWs case, the influence of time difference between LIGO and Fermi is very small, and therefore can be also ignored in the same way.
\section{Constraints on WEP by the maximum time delay}\label{section4}
In this section, we present the constraints on WEP by using the maximum lasting time for the GWs and EM waves signals. According to the reports given by LIGO \cite{LIGO-PRL} and Fermi\cite{VConnaughton}, the waves of GW150914 (or GBM transient 150914) are all over a bandwidth from 35~Hz to 250~Hz (or from 1~keV to 10~MeV). Therefore, two kinds of the waves consist of multi-frequency gravitons or multi-energy photons. For a given frequency of graviton or given energy of photon, there is no definite position at the time axis in the spectrum of time evolution about GW150914 and GBM transient 150914.

Before our calculation, it is necessary to clarify the rationality of the maximum time delay adopted here. For two test particles marked by $1$ and $2$, the time difference due to propagation from the source is $\Delta t1 = t1 (arrive) - t1(emit)$ or $\Delta t2 = t2(arrive) - t2(emit)$. If particles $1$ and $2$ are emitted about in the same time, i.e. $t1(emit) \approx t2(emit)$, the observed time delay is given by
\begin{equation}\label{1-3}
\Delta t_{obs} = |\Delta t1 - \Delta t2| = |t1(arrive) - t2(arrive) |.
\end{equation}
It is just the time difference between particles $1$ and $2$ arrived at detectors, which can be obtained through the spectrum of time evolution. Meanwhile, there is a relationship existed both in GW150914 and GBM transient 150914,
\begin{equation}\label{1-4}
|t1(arrive) - t2(arrive) | \leq \Delta t_{detector},
\end{equation}
where $\Delta t_{detector}$ is the lasting time of signal at detector. To this regard, the maximum delay can be obtained by the way of that particle $1$ is detected at the beginning and particle $2$ is detected at the end. Substituting the delay (\ref{1-4}) into Eq.(\ref{chushi}) with $\Delta \gamma = |\gamma_1 - \gamma_2|$, we can get the difference of $\gamma$ parameters between particles $1$ and $2$,
\begin{equation}\label{4-1-2}
|\gamma_1 - \gamma_2| < \Delta t_{detector} \frac{c^3}{G M_{mw}\log (d/b)},
\end{equation}
where the circumstance of $\Delta t_{gra} < \Delta t_{obs} \leq \Delta t_{detector}$ is considered.

For the frequency-dependent signals of GW150914, the lasting time at detectors is $\Delta t_{detector} = \Delta t_{GW2} \sim 0.45$~s which sweeps upwards from 35~Hz to 250~Hz (see Fig.1 in \cite{LIGO-PRL}). Substituting the coordinate of BFP into Eq.(\ref{4-1-2}), the up-limit of the difference of PPN parameter $\gamma$ for complete signals is shown by
\begin{equation}\label{3-1}
\big|\gamma(35 Hz) - \gamma(250 Hz)\big| < 1.39 ^{+0.08}_{-0.04} \times 10^{-8},
\end{equation}
where $\Delta t_{int}$ is positive if high energy particles arrive later than low energy case \cite{HGao,JJWei}. Like former cases the results of $11$ locations are also illustrated in Fig.\ref{Fig1}. The average value is $1.41 \times 10^{-8}$. It is an order of magnitude looser than the results given by \cite{XFWu} and \cite{EOKahya}. The reason is that the lasting time 0.2~s of inspiral part is about half of the whole swept upward time 0.45~s.
\begin{figure}
\includegraphics[width=3.4 in]{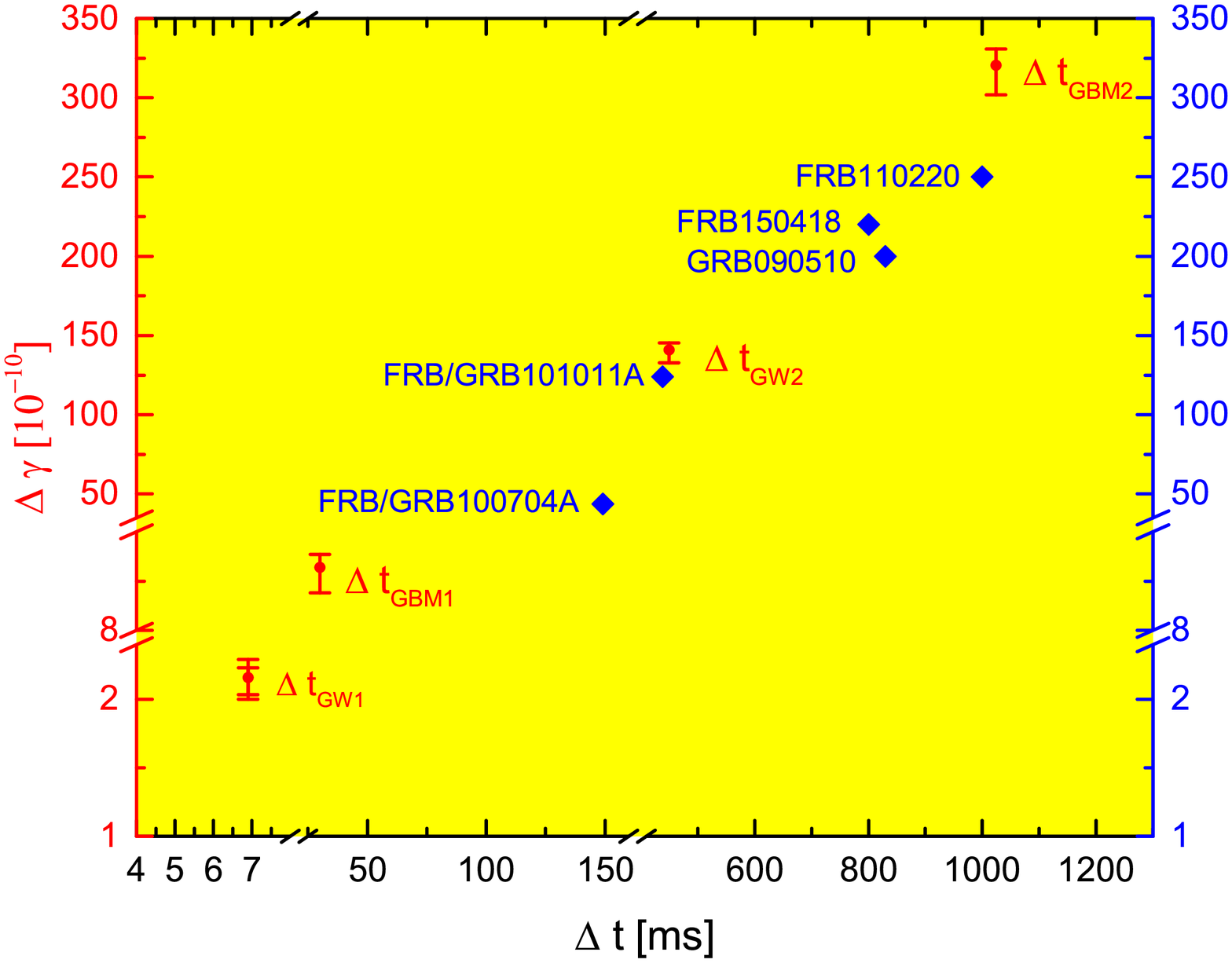}
\caption{The comparison of the up-limit of $\Delta \gamma$ between GW event GW150914 and FRB/GRB. The red circle point denotes our results by adopting the coordinate of BFP. The error bars comes from the uncertainty of luminosity distance $410_{-180}^{+160}$~Mpc provided by LIGO. Only the initial gravitons need to consider the time uncertainty of $6.9_{-0.4}^{+0.5}$~ms. The blue diamond point denotes previous observations of FRB and GRB \cite{HGao,JJWei,SJTingay}.}\label{Fig4}
\end{figure}

For the energy-dependent signals of GBM transient 150914, different energy photons appear in the eight energy channels and the photons with energy from 1~keV to 10~MeV are mixed together (see Fig.8 in \cite{VConnaughton}). As mentioned before, the lasting time of GBM transient 150914 at detectors is $\Delta t_{detector} = \Delta t_{GBM2} \sim 1.024$~s. In the same way, according to the coordinate of BFP and Eq.(\ref{4-1-2}), the up-limit of $\Delta \gamma$ for complete signals is also can be given by
\begin{equation}\label{4-1}
\big|\gamma(1~\text{keV}) - \gamma(10~\text{MeV})\big| < 3.17_{-0.09}^{+0.18} \times 10^{-8}.
\end{equation}
The results of other Fermi's $11$ possible sky locations are drawn in Fig.\ref{Fig1} with red up triangle point. The average value of $\Delta \gamma$ is $3.21 \times 10^{-8}$. Again, all these results about the photons in GBM transient 150914 show the same order of magnitude $10^{-8}$.

At the end of this section, we compare our four kinds of the results with other results that appear in the literatures \cite{JJWei,SJTingay,JJWei2,MJLongo,LMKrauss}. Here, our results are listed by Eqs.(\ref{1-1}), (\ref{2-1}), (\ref{3-1}) and (\ref{4-1}). The compare results are illustrated in Fig.\ref{Fig4}. The result of the initial gravitons with $\Delta t_{GW1}$ is $\Delta \gamma \lesssim 10^{-10}$, which is at least two orders of magnitude tighter than GHz photons from $FRB110220$ \cite{JJWei} and $FRB150418$ \cite{SJTingay}. It is also about one order tighter than GHz photons from $FRB/GRB100704A$.

Fig.\ref{Fig4} also shows that the result $\Delta \gamma \lesssim 10^{-10}$ for $\Delta t_{GBM1}$ is five orders of magnitude tighter than PeV neutrino from blazars \cite{JJWei2}, and is also seven orders of magnitude tighter than MeV neutrino from supernova $1987A$ \cite{MJLongo,LMKrauss}. They are all belong to the test of WEP, by comparing photon with other kinds of particle. We can find the case of graviton is more tighter than the case of neutrino. Hence, the observations of GWs and EM waves could offer us a tighter constraint on WEP. Other two kinds of the results are $\Delta \gamma \lesssim 10^{-8}$ for $\Delta t_{GW2}$ and $\Delta \gamma \lesssim 10^{-8}$ for $\Delta t_{GBM2}$.
\section{Conclusion}\label{section5}
We have tested WEP with GW150914 and the possible EM counterpart. The tests consist of four kinds of comparisons: (a) the initial GWs detected by the two LIGO sites (Hanford and Livingston); (b) the GW signal and the EM signal; (c) the GWs with different frequencies from 35 Hz to 250 Hz; (d) the EM photons with different energies from 1 keV to 10 MeV. Our results show that multi-messenger signals exploiting different emission channels can strengthen existing tests of the WEP.

About the rationality of using initial GWs, there are three points need to be noticed. (i) According to wave-particle duality, initial GWs with given frequency corresponds to initial gravitons with given energy. The latter can be clearly treated as a test particle to constrain WEP. (ii) Based on the data released by LIGO \cite{LIGO_data}, one can find there is a strain gap $\Delta h = h_{L1} - h_{H1} \neq 0$ in GW150914. It means that initial GWs or gravitons received by $L1$ and $H1$ have different polarization and different 4-momentum. Thus we can say initial GWs or gravitons have different internal structure and can be used to test WEP. (iii) Although $H1$ and $L1$ are located at different positions, the difference of their coordinates can be ignored for the cosmic source GW150914 in celestial equatorial coordinate system. The maximum deviation caused by $\Delta L$ is very small to $10^{-19}$, and thus can be neglected.

Except that Fermi claimed they found GBM transient 150914, there was no clear detection by other observation stations such as the INTEGRAL/SPI-ACS \cite{Savchenko}. Therefore, the positive \cite{KMurase,ALoeb,RPerna,BZhang} and the negative \cite{XiongS} opinions were raised to debate this association. For instance in \cite{XiongS}, detectors' viewing angles of GBM transient 150914 was queried. However, up to now this association can not be ruled out by current observation. In order to relieve the impact of the association we choose as many locations as possible. The locations contain one BFP (see Fig.\ref{Fig4}), and 11 positions divided in the LIGO arc (see Fig.\ref{Fig1}). We find different locations actually give the same order of magnitude about $\Delta \gamma$. This point is also confirmed by the whole LIGO arc without considering GBM transient 150914 (see Fig.\ref{Fig2}). But in the comparison between GWs and EM waves in section \ref{section3}, our results are largely impacted by this association because the delay time caused by gravity depends on the radiation model of GBM transient 150914. According to these situations, we can see that our results are more dependent on the distance and delay time than on localization. The distance and delay time can be given out exactly by GW150914. Therefore, even without this association, the results are still partially reliable, at least on the order of magnitude.

In view of the current observation status of the EM counterpart of GWs, it is difficult to confirm the real radiation model including the exact activity time of central engine. The rationality of the parameters $R_8$ and $m_{30}$ in Eq.(\ref{4th-inspiral}) is supported by the condition (\ref{2th-ad-R-m}). In principle, if the condition (\ref{2th-ad-R-m}) is respected well, the parameters $R_8$ and $m_{30}$ are all allowed. Without losing the generality, we choose $R_8 \sim 1$ and $m_{30} \sim 1$ to present a reasonable estimate of intrinsic time delay $\Delta t_{int}$ between GWs and EM waves. Meanwhile, the uncertainty caused by time difference between GWs and EM waves is presented as followings. Comparing GW150914 with GBM transient 150914, one can find the time delay caused by gravity is $\Delta t_{gra} = 0.4~\text{s} - \Delta t_{int}$. In the post-merger emission model there is a relationship $\Delta t_{int} \sim t_{GW}$. In this way, the uncertainty of the time difference between GWs and EM waves can be obtained by the time difference of physical emission of EM waves. The signal GW150914 is dominated by several cycles of a wave pattern whose amplitude is initially increasing. The initial increasing is starting from around the time 0.3~s \cite{LIGO-PRL,LIGO-August}. The entire visible part of the signal lasts around the time of $0.45~\text{s} - 0.3~\text{s} = 0.15~\text{s}$. The inspiral timescale $t_{GW}$ must be in the range of [0.3~s, 0.4~s]. Hence, $\Delta t_{gra}$ should be in the range of $0 < \Delta t_{gra} < 0.1 \text{s}$. It just is the uncertainty of gravitational time delay, which can impact the results of the comparison between GWs and EM waves. To this regard, the maximum of $\Delta \gamma$ is $\Delta \gamma \sim 3.1 \times 10^{-9}$.

Unlike the cases of NS-NS or BH-NS merger with rich EM counterparts, BH-BH merger is not expected to have rich surrounding materials to power bright EM counterparts \cite{BZhang}. However, the relativistic jet may possibly be formed in the merger of BH-BH system in the environments pervaded by large ambient magnetic fields or high baryon densities \cite{LIGO-Localization}. Therefore, one may observe a prompt short gamma-ray burst lasting on the order of one second or less. Searching the afterglow localized at the host galaxy is the best way to determine whether GBM transient 150914 is associated with GW150914 or not. For example, recently a possible semi-analytic afterglow model of GBM transient 150914 was proposed by Trans-Relativistic Afterglow Code (TRAC) \cite{BJMorsony}. They assumed that GBM transient 150914 was an off-axis short GRB with a relatively bright jet ($E_{iso} = 10^{52}~\text{erg}$) and its afterglow could be detected at 863.5 MHz. The brightest X-rays re-brighten was about $6.3 \times 10^{-13}~\text{erg}~\text{cm}^{-2}~\text{s}^{-1}$ below the limit of Swift LMC observations \cite{Swift}.

\begin{acknowledgements}
We thank the anonymous referee for helpful corrections, and also thank Tong Liu of Department of Astronomy in Xiamen University for helpful discussions. This work is supported by the National Natural Science Foundation of China under grants 11475143, 11675032 and 11645003, Science and Technology Innovation Talents in Universities of Henan Province under grant 14HASTIT043, the Nanhu Scholars Program for Young Scholars of Xinyang Normal University. This research has made use of data obtained from the LIGO Open Science Center (https://losc.ligo.org), a service of LIGO Laboratory and the LIGO Scientific Collaboration. LIGO is funded by the U.S. National Science Foundation.
\end{acknowledgements}

\end{document}